\documentclass[sigconf,nonacm]{acmart}

\usepackage{amsmath}
\usepackage{amsfonts}
\usepackage{bm}
\usepackage{multirow}
\usepackage{booktabs}
\usepackage{enumitem}
\usepackage{algorithm}
\usepackage{algpseudocode}
\usepackage{pifont}
\usepackage{colortbl}
\usepackage{makecell}
\usepackage{marvosym}

\definecolor{ourscol}{RGB}{229,89,52}

\newcommand{\ie}{\textit{i.e.}}
\newcommand{\eg}{\textit{e.g.}}

\newcommand{\cmark}{\textcolor[RGB]{214,69,65}{\ding{51}}}
\newcommand{\xmark}{\textcolor[RGB]{54,144,136}{\ding{55}}}

\AtBeginDocument{%
}

\setcopyright{none}
\copyrightyear{2026}
\acmYear{2026}
\renewcommand\footnotetextcopyrightpermission[1]{}
\hypersetup{pdfauthor={Jiakai Tang, Yang Zhang, See-Kiong Ng, Xu Chen, Wen Chen, Jian Wu, Han Zhu},
            pdftitle={Learning from the Future: Privileged Self-Distillation for Sequential Recommendation}}

\begin{document}

\title{Learning from the Future: Privileged Self-Distillation for Sequential Recommendation}

\author{Jiakai Tang$^{1,3}$, Yang Zhang$^{2}$, See-Kiong Ng$^{2}$, Xu Chen$^{1}$\textsuperscript{\,\Letter}, Wen Chen$^{3}$\textsuperscript{\,\Letter}, Jian Wu$^{3}$, Han Zhu$^{3}$}
\affiliation{%
  \institution{$^{1}$Gaoling School of Artificial Intelligence, Renmin University of China, Beijing, China\\
  $^{2}$National University of Singapore, Singapore\\
  $^{3}$Alibaba Group, Beijing, China\\
  tangjiakai5704@ruc.edu.cn}
  \country{}}

\renewcommand{\shortauthors}{Jiakai Tang et al.}

\begin{abstract}
Sequential recommenders are commonly trained with one-hot next-item labels under a causal (\textit{i.e.,} prefix-only) objective aligned with inference. Although deployment-compatible, this supervision provides little information about the relative preferences among non-target items. Yet logged interaction sequences contain an additional source of supervision: interactions following the target often reveal how the user’s intent evolves and make the target easier to interpret. We view these future interactions as training-only privileged information—available during learning but unavailable at inference. This raises a natural question: can future interactions serve as a richer supervisory signal while keeping training aligned with how the model makes predictions at inference?

We propose \textbf{\textit{Privileged Self-Distillation (PSD)}}, a simple framework that separates the information used for learning from the information available at inference. PSD applies two attention masks to the same sequential backbone: a future-aware view produces a privileged \textbf{teacher} distribution conditioned on past and future interactions, while a prefix-only view produces the \textbf{student} distribution used for deployment. Distilling the privileged distribution converts future interactions into training-only supervision rather than inference-time inputs. Because the two views share a backbone, the teacher’s advantage is informational rather than architectural, eliminating the need for a separately pretrained teacher and allowing its supervision to adapt as the student evolves. PSD further employs an advantage-reachability gate to focus distillation on teacher signals more likely to be supported by the observed prefix, together with a momentum-averaged teacher that stabilizes distillation targets. The entire framework is optimized end-to-end in a single stage, leaving the deployed model and its inference cost unchanged. Extensive experiments across public benchmarks and diverse sequential backbones demonstrate consistent and substantial improvements over strong baselines. 
\end{abstract}




\maketitle
\hypersetup{pdfauthor={Jiakai Tang, Yang Zhang, See-Kiong Ng, Xu Chen, Wen Chen, Jian Wu, Han Zhu}}
{\renewcommand{\thefootnote}{}\footnotetext{\Letter\ Co-corresponding authors.}\addtocounter{footnote}{-1}}

\section{Introduction}
Sequential recommendation predicts a user's next interaction from her historical behavior~\cite{boka2024survey,wang2019sequential,pan2026survey}. In deployment, the recommender only observes the user's past interactions and ranks candidate items for the next step. To match this deployment condition, modern sequential recommenders commonly adopt next-item prediction, where a causal model estimates the next-item distribution from the observed prefix and is optimized by cross-entropy on the observed item. This hard-label supervision is serving-compatible but low-information~\cite{szegedy2016rethinking}: it over-confidently assigns all probability mass to the observed target, while leaving the relative preferences among non-target candidates largely under-constrained. As a result, each training position provides only a one-hot target, and the rest of the logged trajectory is left unused as supervision during training.

This unused portion of the trajectory, however, carries exploitable signal in its own right. At training time, the future suffix of the same sequence is also observed, and these future interactions often reflect the underlying evolving user intent that makes the current target easier to interpret, thereby reducing uncertainty about the observed next item and providing supervision beyond the one-hot label. In this sense, future interactions act as \emph{training-only privileged information}~\citep{vapnik2009lupi,lopezpaz2016unifying}: informative during learning but unavailable during serving. This creates a useful but delicate opportunity: future interactions can enrich the training signal, yet the deployed recommender must remain strictly causal. The central question is \textbf{how to learn from the future without serving with the future}.

\begin{figure}[t]
\centering
\includegraphics[width=\linewidth]{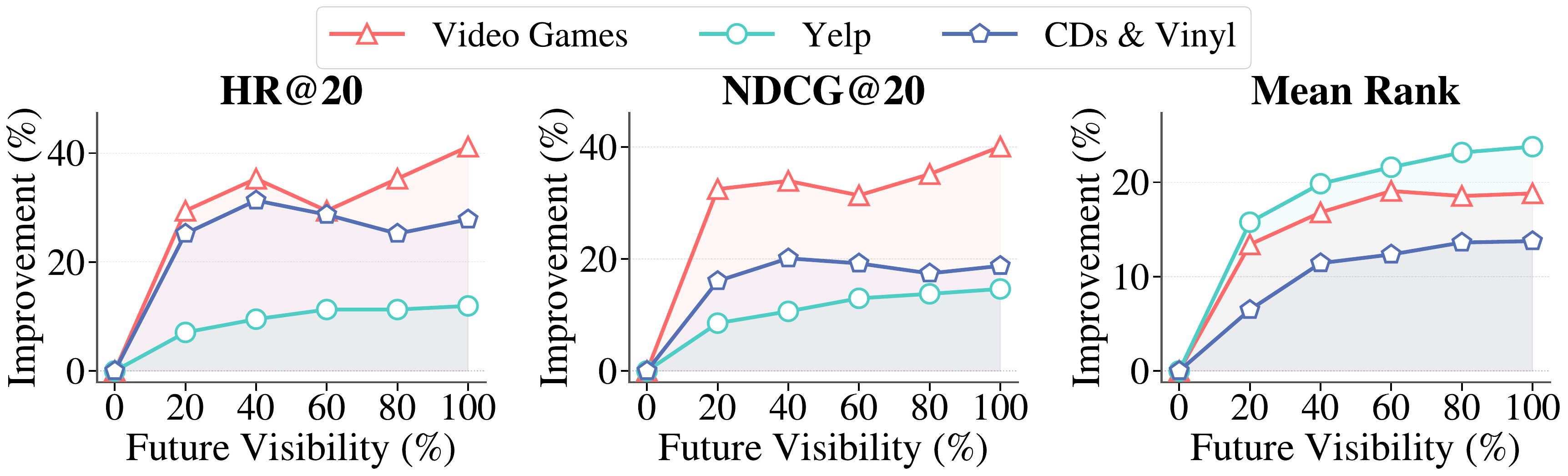}
\vspace{-15pt}
\caption{Relative improvement in HR@20, NDCG@20, and the mean rank of the target item as the visible fraction of future interactions increases (0\% $\to$ 100\%). Even partial future visibility yields substantial gains, indicating that future interactions carry predictive signal well beyond what the causal model extracts from the prefix alone.}
\label{fig:future-visibility}
\vspace{-10pt}
\end{figure}

A seemingly direct solution is to let the model attend to both past and future interactions during training. Bidirectional objectives such as masked-item prediction in BERT4Rec~\citep{bert4rec} follow this intuition and can expose richer context than purely causal next-item prediction. Yet this changes the intrinsic learning problem from causal next-item prediction to reconstruction with unavailable context. The resulting model may learn useful correlations, but the prediction path no longer matches the one used at inference. Figure~\ref{fig:future-visibility} provides a diagnostic view of this tension: future visibility can improve ranking quality, but it also opens a gap between what a privileged model can infer and what a causal model can access. This calls for a mechanism that extracts training signal from future context while keeping the deployed predictor strictly causal.

Knowledge distillation~\cite{gou2021knowledge,xu2024survey,mansourian2025comprehensive} provides a natural way to separate these two roles: a privileged teacher can learn from richer training information and transfer soft-label supervision to a deployable student. However, conventional distillation does not directly address the constraints introduced by future-context privileged information. Some methods rely on separately trained teachers or compact-model transfer pipelines~\citep{tang2018ranking}; others draw supervision from additional label sources, side information, or privileged features collected outside the interaction sequence itself~\citep{wu2024learning,wei2024leave,xu2019privileged}. These choices often introduce additional parameters, auxiliary features, or multi-stage training pipelines, and may create a mismatch between a static teacher and an evolving student: the teacher is frozen or optimized independently, while the student continues to update. Recent studies also suggest that an initially strong but fixed teacher is not always the best supervisor for a weaker student~\citep{agarwal2024onpolicy,gu2024minillm}.

These limitations suggest an alternative approach: rather than importing a stronger teacher, the same model can be exposed to two views of each logged sequence. One view uses the future suffix to form a privileged training signal, while the other remains causal and matches the information path used at inference. This keeps the teacher's advantage within the sequence itself and avoids extra parameters, side features, or a separate teacher-training stage. Coupling the two views, however, raises two nontrivial issues. \textbf{(1)} The privileged teacher encodes much of its extra insight as soft preferences over non-target items, a form of \textbf{\emph{dark knowledge}} for the student, yet not all of it is transferable to the causal student: some preferences are justified only by the future suffix, and forcing the causal student to imitate this unreachable part asks it to explain choices with evidence it will never observe at serving time. \textbf{(2)} The teacher is not an independent source of supervision: it shares its still-learning parameters with the student, so its soft targets are regenerated at every step from an immature model, and its errors are written back into the shared weights through distillation, only to reshape the next round of targets. Without a stable reference, mistakes are self-confirmed rather than corrected, and the privileged signal can be drowned in this feedback loop.

To address these issues, we propose \textbf{Privileged Self-Distillation (PSD)}, a \emph{dual-role in one model} framework built on a single Transformer. PSD evaluates the same backbone under two attention masks: a bidirectional mask produces a privileged teacher distribution conditioned on both past and future interactions, while a causal mask produces the student distribution used at inference. The privileged distribution is transferred to the causal student through teacher-to-student KL distillation, but not all teacher signals are treated equally. PSD applies an \emph{advantage-reachability gate} that discards high-discrepancy distillation terms within each mini-batch, focusing learning on privileged signals that are more likely to be reachable from the causal prefix. To stabilize this self-referential supervision, PSD further maintains a \emph{momentum-averaged teacher}, whose slowly evolving parameters average out transient errors instead of reinforcing them.

Our main contributions are summarized as follows:
\begin{itemize}[leftmargin=*]
    \item We identify future interactions as \emph{training-only privileged information} for sequential recommenders, revealing supervision beyond low-information one-hot next-item labels while preserving the causal deployment costraint.
    \item We propose Privileged Self-Distillation, a simple yet effective framework that constructs a privileged teacher and a causal student through attention-mask control, without requiring extra parameters, auxiliary features, or multi-stage training.
    \item We provide extensive empirical evidence across public benchmarks, diverse sequential backbones, and diagnostic analyses, showing that PSD delivers consistent and substantial gains.
\end{itemize}

\section{Preliminary}

\subsection{Task Formulation}
In sequential recommendation, each user leaves a chronologically ordered interaction sequence $\mathcal{S}=(i_1,i_2,\ldots,i_L)$, where each item $i_t$ belongs to the item set $\mathcal{I}$. Given the prefix $\mathcal{S}_{<t}=(i_1,\ldots,i_{t-1})$, the task is to identify the item that the user interacts with at position $t$. A recommender with parameters $\theta$ accordingly ranks all candidates through a softmax over the item set $\mathcal{I}$:
\begin{equation}
    p_\theta(\cdot \mid \mathcal{S}_{<t}) = \operatorname{softmax}\!\big(\mathbf{z}_{\theta,t}\big),
\end{equation}
where the logits $\mathbf{z}_{\theta,t}$ are produced under a \emph{causal} attention mask, which restricts every position to attend only to itself and its predecessors, so that $\mathbf{z}_{\theta,t}$ is a function of the prefix $\mathcal{S}_{<t}$ alone. The standard learning objective is the next-item cross-entropy over the corpus $\mathcal{D}$ of logged sequences, which is defined as:
\begin{equation}
\label{eq:ce}
\mathcal{L}_{\mathrm{CE}}(\theta)
= -\frac{1}{|\mathcal{D}|}\sum_{\mathcal{S}\in\mathcal{D}}\sum_{t=1}^{|\mathcal{S}|}
\log p_{\theta}\!\left(i_t \mid \mathcal{S}_{<t}\right).
\end{equation}
Minimizing Eq.~\eqref{eq:ce} keeps training aligned with practical serving, since the deployed model likewise ranks candidates from the prefix alone. The price of this alignment lies in the supervision itself: a one-hot label speaks only for the observed item and is silent on all others, so each training position sharpens the likelihood of a single target while leaving the ordering of the remaining catalog essentially unconstrained. The model is thus asked to produce a full ranking at serving time, yet never told during training what a good ranking beyond the observed item looks like.

\begin{figure*}[t]
\centering
\includegraphics[width=\textwidth]{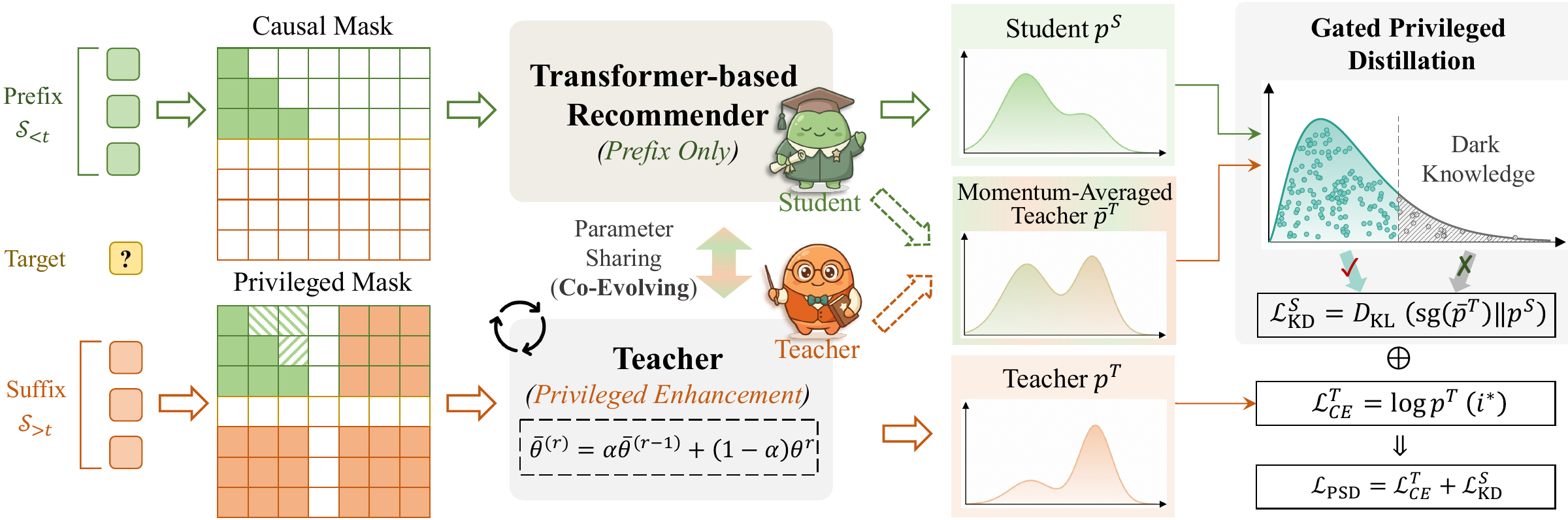}
\vspace{-15pt}
\caption{Overview of PSD. The same Transformer backbone is evaluated under two attention masks: the causal mask restricts the student to the prefix $\mathcal{S}_{<t}$, while the privileged mask additionally exposes the suffix $\mathcal{S}_{>t}$ to the teacher, with all parameters shared and co-evolving. The teacher is grounded on the observed target through $\mathcal{L}_{\mathrm{CE}}^{T}$, and its momentum-averaged distribution $\bar{p}^{T}$ supervises the student through the gated distillation loss $\mathcal{L}_{\mathrm{KD}}^{S}$, which withholds the high-discrepancy dark knowledge that the causal prefix cannot support. Only the causal student is used at inference time.}
\label{fig:psd-framework}
\vspace{-10pt}
\end{figure*}

\subsection{Future as Privileged Information}
A richer signal, however, is already present in the training data but discarded by the causal objective. For a target at position $t$, the suffix $\mathcal{S}_{>t}=(i_{t+1},\ldots,i_L)$ is fully observed during training, yet never available at serving time. This is precisely the setting of \emph{privileged information}~\citep{vapnik2009lupi,lopezpaz2016unifying}: an auxiliary view that may inform learning but must not enter the deployed predictor. The suffix is informative because user behavior is temporally coherent: the interactions that follow $i_t$ record how the user's interests unfold, offering retrospective context that narrows down the plausible targets beyond what the prefix reveals. Formally, the uncertainty it removes about the target is the conditional mutual information
\begin{equation}
\label{eq:cmi}
    I(i_t;\mathcal{S}_{>t}\mid\mathcal{S}_{<t})
    = H(i_t\mid\mathcal{S}_{<t}) - H(i_t\mid\mathcal{S}_{<t},\mathcal{S}_{>t}) \;\geq\; 0,
\end{equation}
which is strictly positive whenever the future is predictive of the target given the past. A predictor conditioned on both sides therefore attains a lower-entropy, and hence lower Bayes-risk, view of the label than any predictor restricted to the prefix.

This asymmetry cannot be closed at inference, since $\mathcal{S}_{>t}$ does not exist once the model is deployed. It can nonetheless be exploited during training, provided the future is used to \emph{shape the learning target} rather than to feed the prediction path. We formalize this through two conditional distributions produced by the same recommender under different information sets: a privileged \emph{teacher}
\begin{equation}
    p_\theta^{T}(\cdot\mid\mathcal{S}_{<t},\mathcal{S}_{>t}),
\end{equation}
which reads both the prefix and the suffix, and a causal \emph{student}
\begin{equation}
    p_\theta^{S}(\cdot\mid\mathcal{S}_{<t}),
\end{equation}
which is restricted to the prefix and is the only distribution used at serving time. The teacher's advantage over the student is exactly the information quantified in Eq.~\eqref{eq:cmi}. This gap, however, is a double-edged signal: part of the teacher's confidence is grounded in patterns the student can recover from the prefix, while part rests solely on the future suffix and is inherently unreachable at inference. The central question, addressed in the next section, is how to transfer the reachable part of this privileged signal to the causal student while suppressing the part that the prefix can never justify.

\section{Methodology}

The architecture of PSD is summarized in Figure~\ref{fig:psd-framework}. PSD builds on a simple observation: a privileged teacher need not be a second model. Evaluated under two attention masks, a single sequential backbone serves both roles, where a \emph{privileged} mask exposes the future suffix to the teacher during training, while a \emph{causal} mask restricts the student to the information path used at serving. Since the two views share all embeddings, Transformer layers, and the prediction head, the teacher's advantage stems purely from richer information access rather than from extra capacity. Upon this construction, two mechanisms address the issues identified in the introduction: an advantage-reachability gate confines distillation to signals the causal prefix can support (Section~\ref{sec:gate}), and a momentum-averaged teacher keeps the self-referential supervision stable (Section~\ref{sec:ema}).

\subsection{Dual-Mask Teacher and Student} 
The teacher and the student are distinguished solely by their attention scope. At a training position $t$, the student mirrors the deployment condition and attends only to the prefix:
\begin{equation}
\label{eq:student-mask}
    \mathcal{A}^{S}_t = \{k \mid k<t\}.
\end{equation}
The teacher adopts a training-time receptive field that additionally covers the suffix:
\begin{equation}
\label{eq:teacher-mask}
    \mathcal{A}^{T}_t = \mathcal{R}_t \cup \{k \mid k>t\},
\end{equation}
where $\mathcal{R}_t\subseteq\mathcal{A}^{S}_t$ is the visibility pattern inside the prefix inherited from the backbone: it remains causal for autoregressive backbones such as SASRec, and could be fully bidirectional for backbones such as BERT4Rec. In either case, the target item $i_t$ is excluded from its own receptive field to prevent label leakage, and the student never observes the suffix $\mathcal{S}_{>t}$, so that the deployed model remains strictly causal. The two receptive fields translate into binary attention masks $\mathbf{M}^{S},\mathbf{M}^{T}\in\{0,1\}^{L\times L}$, where $\mathbf{M}_{t,k}=1$ if and only if position $t$ may attend to position $k$, \ie, $k\in\mathcal{A}_t$.

Let $f_\theta(\mathcal{S};\mathbf{M})$ denote the backbone evaluated under attention mask $\mathbf{M}$. One forward pass per mask yields
\begin{equation}
\label{eq:dual-dist}
\begin{aligned}
    \mathbf{h}_{t}^{S} &= f_\theta(\mathcal{S};\mathbf{M}^{S})_t,
    & p_{t}^{S} &= \operatorname{softmax}(\mathbf{W}\mathbf{h}_{t}^{S}), \\
    \mathbf{h}_{t}^{T} &= f_\theta(\mathcal{S};\mathbf{M}^{T})_t,
    & p_{t}^{T} &= \operatorname{softmax}(\mathbf{W}\mathbf{h}_{t}^{T}),
\end{aligned}
\end{equation}
with a shared output projection $\mathbf{W}$. The two views are supervised asymmetrically. The teacher, which observes the target from both sides, is grounded on the observed item through the next-item cross-entropy objective, which is formally defined as:
\begin{equation}
\label{eq:teacher-ce}
    \mathcal{L}_{\mathrm{CE}}^{T}
    = - \frac{1}{|\mathcal{B}|}\sum_{(\mathcal{S},t)\in\mathcal{B}} \log p_t^{T}(i_t),
\end{equation}
over a mini-batch $\mathcal{B}$ of sequence-position pairs. The student, in contrast, receives \emph{no} direct one-hot supervision: rather than fitting the same hard label from a strictly weaker information set, it is trained to match the teacher's predictive distribution. Specifically, its objective at position $t$ is the teacher-to-student KL divergence
\begin{equation}
\label{eq:student-kd}
    d_t = D_{\mathrm{KL}}\left(\operatorname{sg}(p_{t}^{T})\;\middle\|\;p_t^{S}\right),
\end{equation}
where $\operatorname{sg}(\cdot)$ stops gradients from flowing into the teacher target, so privileged knowledge is transferred in one direction only. This division of labor allows the privileged view to absorb the observed target, while the causal view acquires the softened preference structure that the target alone cannot convey.

\subsection{Gated Privileged Distillation}
\label{sec:gate}
Through Eq.~\eqref{eq:student-kd}, the student inherits from the teacher a preference structure over the entire item set, a strictly richer supervisory signal than the one-hot label. Not every position, however, warrants this transfer. The divergence $d_t$ itself provides a natural selection criterion, as it measures the discrepancy between the student's causal prediction and the teacher's privileged one at each position. If every teacher preference were recoverable from the prefix, uniformly averaging $d_t$ over the batch would suffice. In practice, however, a large $d_t$ may arise for two reasons: \ding{182} the teacher may rely on evidence available only in the suffix, or \ding{183} the pattern may exceed what the student can absorb at its current stage of training. In either case, enforcing an immediate match imposes supervision that the student cannot yet justify, turning the privileged signal into a source of interference rather than guidance.

To address the above problem, PSD introduces an \textbf{\emph{advantage-reachability gate}} that restricts distillation to positions whose privileged signals are transferable. Given a mini-batch $\mathcal{B}$, let
\begin{equation}
    \gamma_\delta = \operatorname{Percentile}_{\delta}\left(\{d_t\mid(\mathcal{S},t)\in\mathcal{B}\}\right)
\end{equation}
denote the $\delta$-th percentile of the discrepancies within the batch, and define the gating indicator of each position as
\begin{equation}
\label{eq:gate}
    g_t = \mathbb{I}(d_t \leq \gamma_\delta).
\end{equation}
The distillation loss is computed over the admitted positions only:
\begin{equation}
\label{eq:kd-loss}
    \mathcal{L}_{\mathrm{KD}}^{S}
    = \frac{\sum_{(\mathcal{S},t)\in\mathcal{B}} g_t \, d_t}
    {\sum_{(\mathcal{S},t)\in\mathcal{B}} g_t}.
\end{equation}
Positions rejected by the gate are not discarded from learning: they still ground the teacher through $\mathcal{L}_{\mathrm{CE}}^{T}$, but their soft targets are withheld from the student. The gate is named accordingly: a small $d_t$ indicates a teacher advantage that is already reachable from the prefix, whereas a large $d_t$ corresponds to an advantage that is not yet transferable, either because it is grounded solely in the suffix or because it exceeds the student's current capability. Moreover, the rejection is not permanent: the percentile $\delta$ controls the fraction of positions distilled in each batch, and since $\gamma_\delta$ is recomputed within every batch, positions withheld early in training may re-enter distillation as the student improves. The criterion thus adapts jointly to the evolving scale of the discrepancies and to the student's growing capability, without relying on a fixed absolute threshold.

\begin{algorithm}[t]
\caption{Training procedure of PSD}
\label{alg:psd}
\begin{algorithmic}[1]
\Require training corpus $\mathcal{D}$; backbone $f_\theta$; gate percentile $\delta$; EMA rate $\alpha$
\State initialize parameters $\theta$; EMA parameters $\bar{\theta}\leftarrow\theta$
\While{not converged}
  \State sample a mini-batch $\mathcal{B}$ of sequence-position pairs $(\mathcal{S},t)$
  \State $p_t^{S}\leftarrow$ forward pass of $f_\theta$ under causal mask $\mathbf{M}^{S}$ \Comment{Eq.~\eqref{eq:dual-dist}}
  \State $p_t^{T}\leftarrow$ forward pass of $f_\theta$ under privileged mask $\mathbf{M}^{T}$
  \State $\bar{p}_t^{T}\leftarrow$ forward pass of $f_{\bar{\theta}}$ under $\mathbf{M}^{T}$, no gradient \Comment{Eq.~\eqref{eq:ema-teacher}}
  \State $d_t\leftarrow D_{\mathrm{KL}}\big(\bar{p}_t^{T}\,\|\,p_t^{S}\big)$ for each $(\mathcal{S},t)\in\mathcal{B}$
  \State $\gamma_\delta\leftarrow$ $\delta$-th percentile of $\{d_t\}$; $\;g_t\leftarrow\mathbb{I}(d_t\le\gamma_\delta)$ \Comment{Eq.~\eqref{eq:gate}}
  \State $\mathcal{L}_{\mathrm{KD}}^{S}\leftarrow \big(\sum_t g_t d_t\big)/\sum_t g_t$
  \State $\mathcal{L}_{\mathrm{PSD}}\leftarrow \mathcal{L}_{\mathrm{CE}}^{T}+\mathcal{L}_{\mathrm{KD}}^{S}$
  \State update $\theta$ by descending $\nabla_\theta \mathcal{L}_{\mathrm{PSD}}$
  \State $\bar{\theta}\leftarrow\alpha\bar{\theta}+(1-\alpha)\theta$ \Comment{Eq.~\eqref{eq:ema}}
\EndWhile
\State \Return causal student $p_t^{S}$ under mask $\mathbf{M}^{S}$ for serving
\end{algorithmic}
\end{algorithm}

\subsection{Momentum-Averaged Teacher}
\label{sec:ema}
The teacher in PSD is not a fixed oracle: it is produced by the same parameters that are still being updated, so its prediction at a single step may encode transient errors of the current model state. To stabilize this self-referential supervision, following prior works~\citep{tarvainen2017mean,he2020moco,caron2021dino}, PSD maintains an \textit{Exponential Moving Average (EMA)} of the model parameters $\theta$ as $\bar{\theta}$. At each training step, the EMA is updated by the following recursive formula:
\begin{equation}
\label{eq:ema}
    \bar{\theta}^{(r)} = \alpha \, \bar{\theta}^{(r-1)} + (1-\alpha) \, \theta^{(r)},
\end{equation}
where $r$ indexes training steps and $\alpha\in[0,1)$ controls the degree of smoothing, with $\bar{\theta}^{(0)}$ initialized to $\theta^{(0)}$. The distillation target is then produced by this averaged model under the privileged mask,
\begin{equation}
\label{eq:ema-teacher}
    \bar{p}_{t}^{T} = \operatorname{softmax}\big(\bar{\mathbf{W}} \, f_{\bar{\theta}}(\mathcal{S};\mathbf{M}^{T})_t\big),
\end{equation}
so Eq.~\eqref{eq:student-kd} is always computed against $\bar{p}_{t}^{T}$ rather than the instantaneous $p_t^{T}$. Since $\bar{\theta}$ aggregates the whole optimization trajectory, momentary errors of any single step are averaged out over time instead of being written back into the shared weights. The averaged parameters are updated without gradients, exist only during training, and leave the inference graph untouched.

\subsection{Optimization Objective}
In this subsection, we introduce the overall training objective and describe how PSD behaves at inference. The model is optimized end-to-end by a single objective that grounds the teacher on the observed target and guides the student through gated distillation:
\begin{equation}
\label{eq:overall}
    \mathcal{L}_{\mathrm{PSD}}
    = \mathcal{L}_{\mathrm{CE}}^{T}
    + \mathcal{L}_{\mathrm{KD}}^{S}.
\end{equation}
Notably, the objective contains \textit{NO} student cross-entropy term: the causal view is shaped entirely by the teacher's softened distribution rather than by the one-hot target. Since both terms act on one set of shared parameters, the teacher and the student are \textbf{\emph{co-evolving}} during training: each update refines the teacher's privileged view, and the refinement is delivered to the student in the next round of distillation, so the guidance keeps pace with the student's competence rather than growing stale as with a frozen teacher.

The whole model is trained in a single stage without any pre-training or teacher-warmup phase; Algorithm~\ref{alg:psd} summarizes the complete procedure. At inference, recommendations are produced solely from the student distribution $p_t^{S}$ under the causal mask. The deployed model and its serving cost are therefore identical to the original backbone, with no extra parameters or additional passes.

\subsection{Discussion}
\label{sec:discussion}
In this section, we discuss the differences between PSD and representative distillation-based methods for recommendation, with the comparison summarized in Table~\ref{tab:method-comparison}. The essential distinction lies in where the teacher's advantage originates. Existing methods import it from outside the deployed model: RD~\citep{tang2018ranking} relies on a separately pre-trained ranking model, PFD~\citep{xu2019privileged} on scenario-specific privileged features whose availability hardly generalizes across different scenarios, and CSRec~\citep{wu2024learning} and S$^4$Rec~\citep{wei2024leave} on auxiliary teacher modules attached to the backbone. Such an external teacher, however, comes at a price in at least one of the following three aspects:
\begin{itemize}[leftmargin=*]
    \item \textbf{Extra parameters:} all compared methods maintain teacher components that the deployed model never uses, whereas the teacher of PSD is induced from the deployed backbone by an attention mask alone, so our self-distillation framework requires no extra parameters beyond the original backbone.
    \item \textbf{Auxiliary features:} PFD requires privileged features to be collected and maintained in production, whereas the privileged signal of our proposed method, the future suffix of the user sequence, is already present in every logged sequence.
    \item \textbf{Training pipeline:} RD pre-trains its teacher before distillation and CSRec involves multi-stage optimization, whereas PSD optimizes its two loss terms jointly in a single stage.
\end{itemize}
These properties amount to more than engineering economy. Since the teacher of PSD shares the parameter space of the student, its guidance is renewed at every update, yielding the co-evolving supervision analyzed above; and since the construction requires nothing beyond an attention mask, PSD applies to any sequential backbone without modification or extra features, making it a general-purpose training strategy for Transformer-based sequential recommenders.

\begin{table}[t]
\caption{Comparison between PSD and the representative distillation-based training strategies for recommendation.}
\vspace{-10pt}
\label{tab:method-comparison}
\centering
\begin{tabular}{lccc}
\toprule
\textbf{Method} &
\makecell{w/o Extra \\ Parameters} &
\makecell{w/o Auxiliary \\ Features} &
\makecell{One-Stage \\ Training} \\
\midrule
\rowcolor{gray!10}
RD~\citep{tang2018ranking} & \xmark & \cmark & \xmark \\
PFD~\citep{xu2019privileged} & \xmark & \xmark & \cmark \\
\rowcolor{gray!10}
CSRec~\citep{wu2024learning} & \xmark & \cmark & \xmark \\
S$^4$Rec~\citep{wei2024leave} & \xmark & \cmark & \cmark \\
\rowcolor{ourscol!12}
\textbf{PSD (Ours)} & \cmark & \cmark & \cmark \\
\bottomrule
\end{tabular}
\vspace{-10pt}
\end{table}

\section{Experiments}

\subsection{Experimental Setup}

\subsubsection{\textbf{Datasets.}}
We conduct experiments on three widely used benchmarks: two subsets of the Amazon review corpus~\cite{hou2026bridging}, \textbf{Video Games} and \textbf{CDs \& Vinyl}, and the \textbf{Yelp} business-review dataset. On the Amazon subsets, we discard users with fewer than five interactions; on Yelp, we apply the standard 5-core preprocessing so that every user and every item retains at least five interactions. Across all datasets, an interaction is regarded as positive only if its rating exceeds 3, and the remaining interactions are removed. To avoid temporal leakage, all splits follow absolute timestamps: the Amazon subsets adopt the officially released train/validation/test partition\footnote{\url{https://amazon-reviews-2023.github.io/data_processing/5core.html}}, and Yelp is split chronologically with a ratio of 8:1:1. For text-based recommenders, we retain the title, price, and description of each Amazon item, and the name, category, city, and state of each Yelp business, and encode the text with Qwen3-Embedding-8B~\cite{qwen3embedding}. Table~\ref{tab:dataset-statistics} summarizes the statistics of the processed datasets.

\subsubsection{\textbf{Baseline Methods.}}
To examine the generality of PSD across architectures, we instantiate it on three representative sequential recommendation backbones: SASRec~\citep{sasrec}, a unidirectional Transformer trained with next-item prediction; BERT4Rec~\citep{bert4rec}, a bidirectional Transformer trained with masked-item prediction; and UniSRec~\cite{hou2022towards}, a text-enhanced recommender built on pre-trained language representations. On each backbone, \textbf{PSD} is compared with the vanilla training recipe (denoted \textbf{Base}) and several closely related training strategies: Ranking Distillation (\textbf{RD})~\citep{tang2018ranking}, which distills from a separately trained ranking teacher; \textbf{TCE} and \textbf{RCE}~\citep{wang2021denoising}, which truncate or reweight the training loss to resist noisy interactions; \textbf{S$^4$Rec}~\citep{wei2024leave}, an online self-distillation strategy for sequential recommendation; and three variants of CSRec~\citep{wu2024learning}, denoted \textbf{CS-D}, \textbf{CS-M}, and \textbf{CS-T}. All strategies are applied to the same backbone as PSD for a fair comparison, and the hyperparameters of each method are tuned to its best performance on the validation set.

\subsubsection{\textbf{Evaluation Metrics.}}
Following prior works~\cite{tang2026think,tang2026parallel,yuan2023go}, we evaluate top-$K$ recommendation quality with two standard metrics, Hit Ratio (\textbf{HR@$\bm{K}$}) and Normalized Discounted Cumulative Gain (\textbf{NDCG@$\bm{K}$}), reported at $K=10$ and $K=20$. HR@$K$ measures whether the ground-truth item appears in the top-$K$ list, while NDCG@$K$ further rewards placing it at a higher position. We rank the target item against the full item corpus to avoid sampling bias.

\subsubsection{\textbf{Implementation Details.}}
We implement all methods in PyTorch framework. For a fair comparison, the learning rate, batch size, and embedding dimension are fixed to $1\times 10^{-3}$, 2048, and 256, respectively, for all methods. The dropout ratio and the weight decay are tuned over $\{0, 0.1, 0.2, 0.3, 0.4, 0.5\}$ and $\{0, 10^{-5}, 10^{-4}, 10^{-3}\}$, respectively. For PSD, the EMA rate $\alpha$ is searched over $\{0.6, 0.7, 0.8, 0.9, 0.99\}$, and the gate percentile $\delta$ over $\{0.6, 0.7, 0.8, 0.9, 0.99, 1.0\}$, where $\delta=1.0$ disables the gate and distills every position.  We adopt an early stopping strategy that monitors NDCG@20 on the validation set with a patience of 10 epochs. Note that when applying PSD, the student always performs causal inference regardless of the backbone, while the visibility pattern $\mathcal{R}_t$ within the teacher's prefix follows the backbone's own convention, \eg, unidirectional for SASRec and bidirectional for BERT4Rec.

\begin{table}[t]
\caption{Statistics of the benchmark datasets.}
\vspace{-10pt}
\label{tab:dataset-statistics}
\centering
\resizebox{\linewidth}{!}{
\begin{tabular}{lrrrrr}
\toprule
Dataset & \#Users & \#Items & \#Inter. & Avg. Len. & Sparsity \\
\midrule
Video Games & 54,001 & 22,735 & 562,193 & 10.41 & 99.95\% \\
CDs \& Vinyl & 89,590 & 88,462 & 1,347,178 & 15.04 & 99.98\% \\
Yelp & 120,827 & 71,819 & 1,880,704 & 15.57 & 99.98\% \\
\bottomrule
\end{tabular}
}
\vspace{-10pt}
\end{table}

\begin{table*}[t]
\caption{Overall performance on three recommendation benchmarks with different backbone models. The best and second-best results are highlighted in bold and \underline{underlined}, respectively.}
\vspace{-5pt}
\label{tab:overall-results}
\centering
\begin{tabular*}{\textwidth}{@{\extracolsep{\fill}}lllccccccccc}
\toprule
Dataset & Backbone & Metric & Base & RD & TCE & RCE & S$^4$Rec & CS-D & CS-M & CS-T & PSD \\
\midrule
\multirow{12}{*}{Video Games}
& \multirow{4}{*}{SASRec} & HR@10 & 0.0455 & \underline{0.0509} & 0.0479 & 0.0455 & 0.0464 & 0.0485 & 0.0497 & 0.0458 & \textbf{0.0538} \\
& & HR@20 & 0.0747 & \underline{0.0785} & 0.0741 & 0.0678 & 0.0720 & 0.0735 & 0.0711 & 0.0758 & \textbf{0.0818} \\
& & NDCG@10 & 0.0203 & \underline{0.0230} & 0.0212 & 0.0227 & 0.0222 & 0.0214 & 0.0222 & 0.0209 & \textbf{0.0238} \\
& & NDCG@20 & 0.0276 & \underline{0.0299} & 0.0278 & 0.0283 & 0.0287 & 0.0277 & 0.0276 & 0.0284 & \textbf{0.0308} \\
\cmidrule(lr){2-12}
& \multirow{4}{*}{Bert4Rec} & HR@10 & 0.0390 & \underline{0.0416} & 0.0360 & 0.0348 & 0.0402 & 0.0396 & 0.0402 & 0.0407 & \textbf{0.0476} \\
& & HR@20 & 0.0601 & 0.0631 & 0.0577 & 0.0500 & 0.0619 & \underline{0.0645} & 0.0634 & 0.0634 & \textbf{0.0690} \\
& & NDCG@10 & 0.0211 & 0.0218 & 0.0184 & 0.0184 & 0.0219 & \underline{0.0222} & 0.0215 & 0.0221 & \textbf{0.0250} \\
& & NDCG@20 & 0.0264 & 0.0271 & 0.0239 & 0.0222 & 0.0273 & \underline{0.0285} & 0.0273 & 0.0277 & \textbf{0.0304} \\
\cmidrule(lr){2-12}
& \multirow{4}{*}{UniSRec} & HR@10 & 0.0693 & 0.0595 & 0.0654 & 0.0574 & 0.0675 & \underline{0.0699} & 0.0696 & 0.0660 & \textbf{0.0753} \\
& & HR@20 & 0.1044 & 0.0964 & 0.1014 & 0.0925 & 0.0996 & \underline{0.1068} & 0.1032 & 0.1020 & \textbf{0.1112} \\
& & NDCG@10 & 0.0330 & 0.0288 & 0.0308 & 0.0284 & 0.0316 & \underline{0.0335} & 0.0330 & 0.0307 & \textbf{0.0350} \\
& & NDCG@20 & 0.0419 & 0.0380 & 0.0398 & 0.0372 & 0.0397 & \underline{0.0427} & 0.0415 & 0.0397 & \textbf{0.0440} \\
\midrule
\multirow{12}{*}{CDs \& Vinyl}
& \multirow{4}{*}{SASRec} & HR@10 & 0.0415 & 0.0463 & 0.0431 & 0.0418 & 0.0411 & 0.0438 & \underline{0.0464} & 0.0454 & \textbf{0.0485} \\
& & HR@20 & 0.0582 & \underline{0.0649} & 0.0580 & 0.0560 & 0.0646 & 0.0628 & 0.0631 & 0.0639 & \textbf{0.0697} \\
& & NDCG@10 & 0.0189 & 0.0211 & 0.0196 & 0.0212 & 0.0189 & 0.0208 & \underline{0.0214} & 0.0212 & \textbf{0.0226} \\
& & NDCG@20 & 0.0231 & 0.0258 & 0.0234 & 0.0248 & 0.0249 & 0.0256 & 0.0256 & \underline{0.0259} & \textbf{0.0280} \\
\cmidrule(lr){2-12}
& \multirow{4}{*}{Bert4Rec} & HR@10 & 0.0242 & 0.0253 & 0.0238 & \underline{0.0296} & 0.0293 & 0.0261 & 0.0251 & 0.0232 & \textbf{0.0406} \\
& & HR@20 & 0.0380 & 0.0376 & 0.0379 & \underline{0.0448} & 0.0446 & 0.0395 & 0.0392 & 0.0398 & \textbf{0.0566} \\
& & NDCG@10 & 0.0134 & 0.0132 & 0.0127 & 0.0154 & \underline{0.0157} & 0.0141 & 0.0131 & 0.0114 & \textbf{0.0235} \\
& & NDCG@20 & 0.0168 & 0.0163 & 0.0163 & 0.0192 & \underline{0.0196} & 0.0174 & 0.0167 & 0.0156 & \textbf{0.0275} \\
\cmidrule(lr){2-12}
& \multirow{4}{*}{UniSRec} & HR@10 & 0.0386 & 0.0392 & 0.0386 & 0.0373 & 0.0398 & \underline{0.0424} & 0.0405 & 0.0405 & \textbf{0.0438} \\
& & HR@20 & 0.0633 & 0.0607 & 0.0643 & 0.0588 & \underline{0.0652} & 0.0649 & 0.0637 & 0.0631 & \textbf{0.0666} \\
& & NDCG@10 & 0.0188 & 0.0190 & 0.0194 & 0.0183 & 0.0191 & \underline{0.0202} & 0.0199 & 0.0197 & \textbf{0.0207} \\
& & NDCG@20 & 0.0250 & 0.0245 & \underline{0.0259} & 0.0237 & 0.0255 & 0.0258 & 0.0258 & 0.0255 & \textbf{0.0264} \\
\midrule
\multirow{12}{*}{Yelp}
& \multirow{4}{*}{SASRec} & HR@10 & 0.0556 & 0.0565 & 0.0550 & 0.0551 & 0.0570 & \underline{0.0574} & 0.0574 & 0.0573 & \textbf{0.0608} \\
& & HR@20 & 0.0853 & 0.0870 & 0.0849 & 0.0850 & \underline{0.0902} & 0.0877 & 0.0892 & 0.0885 & \textbf{0.0931} \\
& & NDCG@10 & 0.0311 & 0.0307 & 0.0307 & 0.0303 & 0.0307 & 0.0316 & 0.0312 & \underline{0.0317} & \textbf{0.0338} \\
& & NDCG@20 & 0.0386 & 0.0383 & 0.0382 & 0.0378 & 0.0390 & 0.0392 & 0.0392 & \underline{0.0395} & \textbf{0.0419} \\
\cmidrule(lr){2-12}
& \multirow{4}{*}{Bert4Rec} & HR@10 & 0.0474 & 0.0465 & 0.0457 & 0.0460 & \underline{0.0554} & 0.0483 & 0.0468 & 0.0487 & \textbf{0.0578} \\
& & HR@20 & 0.0769 & 0.0765 & 0.0754 & 0.0755 & \underline{0.0882} & 0.0804 & 0.0788 & 0.0812 & \textbf{0.0917} \\
& & NDCG@10 & 0.0233 & 0.0232 & 0.0224 & 0.0227 & \underline{0.0276} & 0.0241 & 0.0232 & 0.0249 & \textbf{0.0311} \\
& & NDCG@20 & 0.0306 & 0.0307 & 0.0298 & 0.0301 & \underline{0.0358} & 0.0321 & 0.0312 & 0.0330 & \textbf{0.0396} \\
\cmidrule(lr){2-12}
& \multirow{4}{*}{UniSRec} & HR@10 & 0.0596 & 0.0595 & 0.0595 & 0.0600 & \underline{0.0625} & 0.0615 & 0.0620 & 0.0590 & \textbf{0.0679} \\
& & HR@20 & 0.0971 & 0.0951 & 0.0956 & 0.0956 & \underline{0.0982} & 0.0971 & 0.0982 & 0.0968 & \textbf{0.1049} \\
& & NDCG@10 & 0.0300 & 0.0299 & 0.0304 & 0.0306 & \underline{0.0317} & 0.0311 & 0.0314 & 0.0297 & \textbf{0.0352} \\
& & NDCG@20 & 0.0394 & 0.0388 & 0.0395 & 0.0396 & \underline{0.0407} & 0.0401 & 0.0405 & 0.0392 & \textbf{0.0445} \\
\bottomrule
\end{tabular*}
\vspace{-5pt}
\end{table*}

\subsection{Overall Performance}
The overall performance comparison is reported in Table~\ref{tab:overall-results}. From the table, we could observe several key findings as follows:
\begin{itemize}[leftmargin=*]
    \item Among the three backbones, BERT4Rec generally underperforms SASRec despite observing richer bidirectional context during training, trailing by nearly 20\% on average and by up to 41.7\% for HR@10 on CDs \& Vinyl. The reason is that its masked-item objective conditions each prediction on both sides of the target, a context that never exists at serving; future information consumed as \emph{input} thus fails to translate into causal ranking ability, and leaked future signals help little unless they align with the serving mode. On the other hand, UniSRec delivers the strongest base performance among the three backbones, benefiting from the rich semantics carried by pre-trained textual representations.
    \item No baseline strategy is consistently competitive. RD, whose supervision comes from a separately pre-trained static teacher, raises the HR@10 of SASRec from 0.0455 to 0.0509 on Video Games yet drags UniSRec down from 0.0693 to 0.0595, since a frozen teacher offers little headroom once the backbone itself grows stronger. The denoising objectives TCE and RCE rely on fixed loss-reshaping heuristics, and accordingly help only occasionally while often falling below the backbone. S$^4$Rec, equipped with auxiliary self-distillation modules, ranks second mainly on Yelp, where it lifts the NDCG@20 of BERT4Rec from 0.0306 to 0.0358, but this advantage does not extend across the other settings; likewise, none of the three CSRec variants shows a stable advantage over the others. Their benefits thus transfer unreliably across backbones and datasets, not to mention the extra parameters, auxiliary features, or multi-stage pipelines these strategies additionally require, as discussed in Section~\ref{sec:discussion}.
    \item PSD achieves the superior performance across all backbones and datasets, improving over the corresponding backbone by 19.8\% on average and over the strongest baseline in each setting by 9.9\%; unlike the static supervision above, its co-evolving teacher keeps the guidance matched to the student throughout training. The improvements are the largest on BERT4Rec, averaging 35.9\% and peaking at 75.4\% for NDCG@10 on CDs \& Vinyl, where the serving-aligned supervision corrects the most severe mismatch, while on the text-enhanced UniSRec, whose base performance is already strong, PSD still brings average gains from 6.6\% to 13.1\%, indicating that the privileged future signal is complementary to rich item semantics. Notably, PSD shows no single case of regression, since the advantage-reachability gate withholds transfers that the causal view cannot yet absorb, and the momentum-averaged teacher keeps the self-referential supervision stable, so the student is not misled by untransferable guidance.
\end{itemize}

\begin{figure}[t]
\centering
\includegraphics[width=0.95\linewidth]{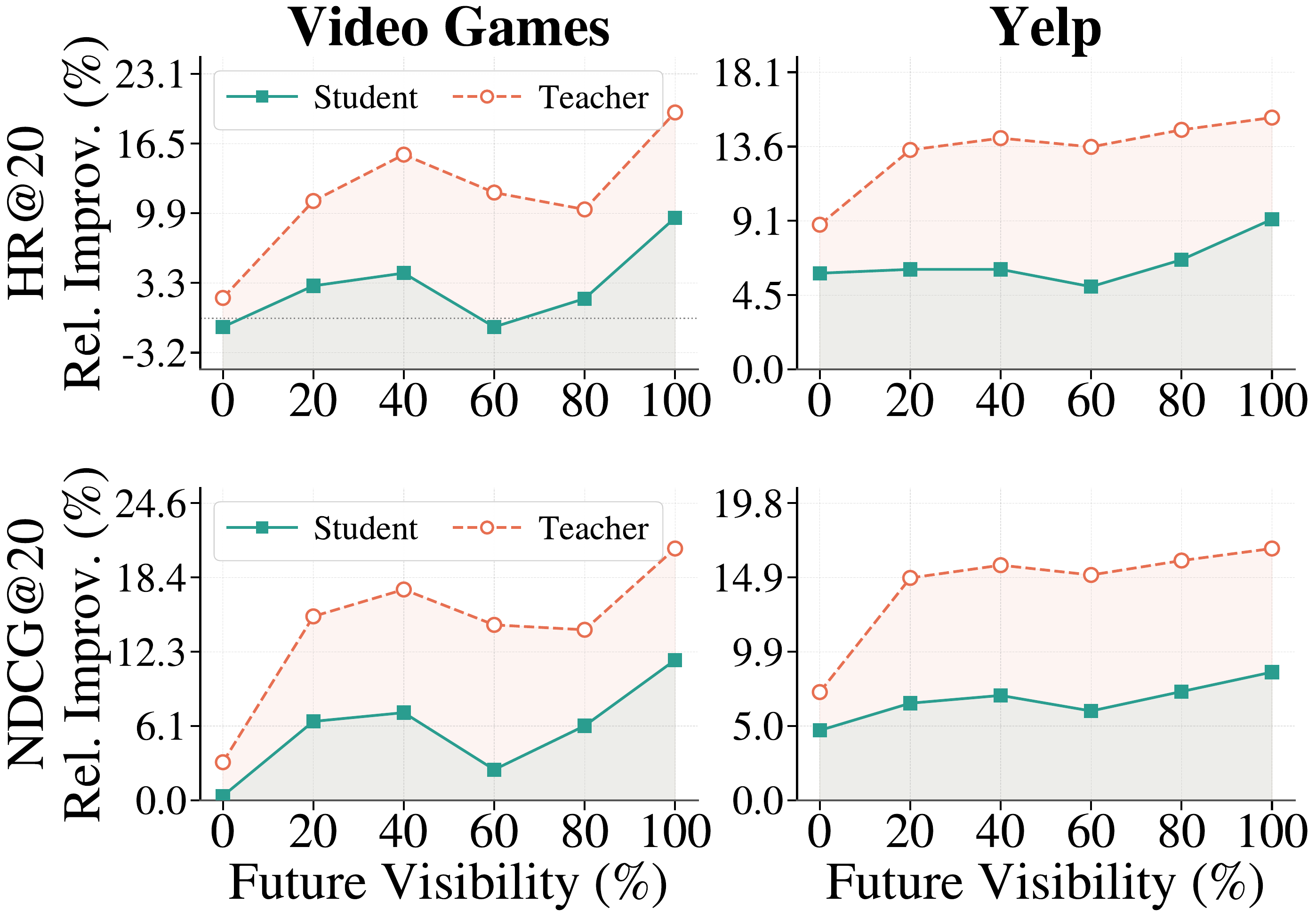}
\vspace{-5pt}
\caption{Relative improvement of the teacher and student over the SASRec backbone under varying future visibility.}
\label{fig:teacher-student}
\vspace{-5pt}
\end{figure}

\subsection{Impact of Future Visibility}
We first examine how much the privileged teacher actually benefits from the future, and how much of this benefit reaches the causal student. On the SASRec backbone, we vary the fraction of the suffix visible to the teacher during training from 0\% to 100\%, and Figure~\ref{fig:teacher-student} presents the relative improvement of the two views over the backbone. The teacher's improvement grows steadily with visibility, reaching roughly 20\% in HR@20 on Video Games at full visibility, which confirms that the suffix carries substantial predictive signal. The student improves in tandem with the teacher, capturing about half of the teacher's gain, and its curve rises and dips at the same visibility levels as the teacher's; this co-movement indicates that the student's gains are indeed inherited from the privileged knowledge. Nevertheless, a clear gap persists at every visibility level and widens as more future context is exposed, indicating that part of the privileged advantage is grounded solely in the suffix and can never be reproduced from the prefix, which is exactly the portion our advantage-reachability gate is designed to withhold. 

\begin{figure}[t]
\centering
\includegraphics[width=\linewidth]{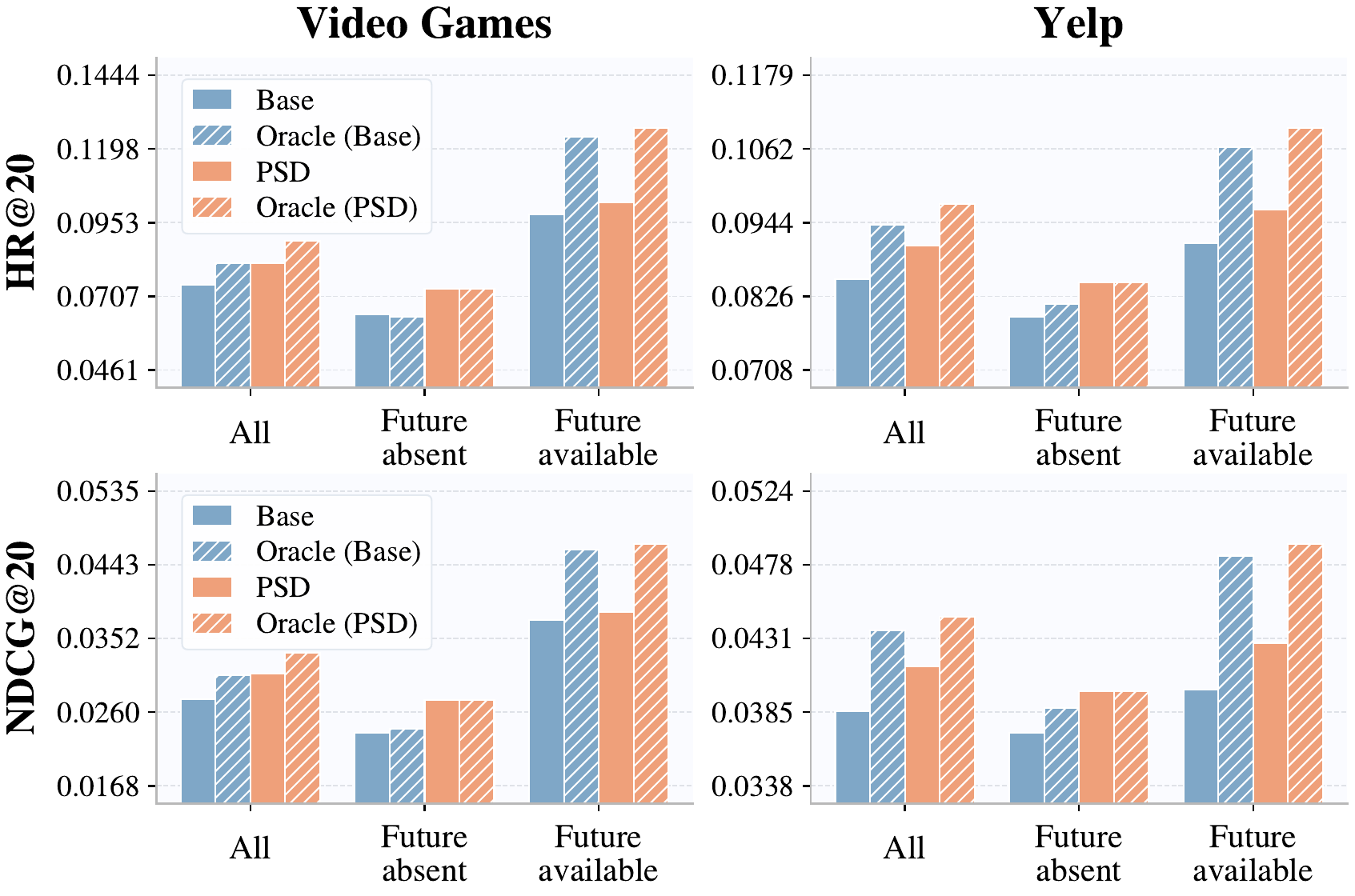}
\vspace{-20pt}
\caption{Oracle analysis comparing PSD with its backbone under different future-information settings.}
\label{fig:oracle}
\vspace{-5pt}
\end{figure}

\subsection{Oracle Analysis}
In this section, we further quantify how much privileged knowledge PSD internalizes into the causal view. For each trained model, we additionally evaluate an \emph{oracle} variant that is granted the privileged mask at test time, and we partition the test cases by whether the target has future interactions recorded in the log. Figure~\ref{fig:oracle} presents the results on Video Games and Yelp with the SASRec backbone, from which three observations emerge as follows:
\begin{enumerate}[wide, labelindent=0pt, label=\textbf{(\roman*)}]
    \item The oracle evaluation lifts the backbone substantially only on the future-available group and leaves the future-absent group unchanged, verifying that the oracle advantage indeed originates from future context rather than from the extra mask itself.
    \item The causal PSD model closes much of this gap without accessing any future input: its performance approaches that of the oracle-evaluated backbone, and it outperforms the backbone even on the future-absent group where no future context exists for any model to exploit, so the distilled preference structure generalizes rather than depending on the availability of the future at test time.
    \item The oracle variant of PSD also surpasses the oracle-evaluated backbone, revealing that the benefit of co-evolution flows in both directions: distillation not only equips the student with privileged knowledge, but the improved student, through the shared parameters, in turn strengthens the privileged view beyond what a separately trained teacher would attain, so the two views reinforce each other through our proposed dual-mode distillation in one model.
\end{enumerate}

\subsection{Further Analysis}

\subsubsection{\textbf{Training Dynamics.}}
\begin{figure}[t]
\centering
\includegraphics[width=\linewidth]{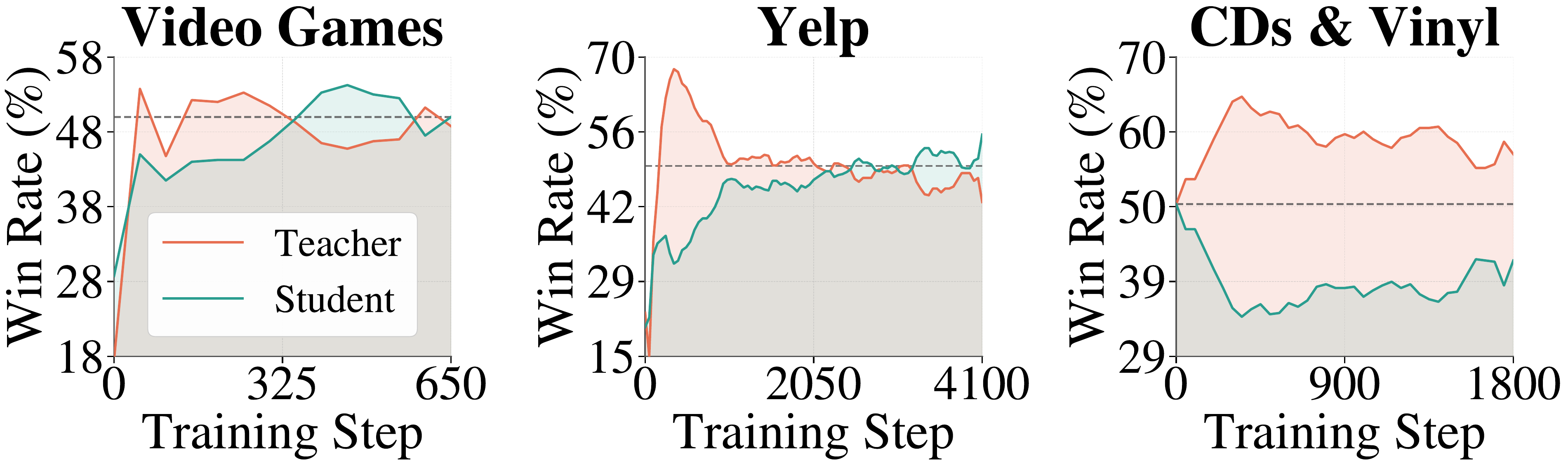}
\vspace{-15pt}
\caption{Win-rate dynamics of the privileged teacher and causal student during training.}
\label{fig:rankgap}
\vspace{-10pt}
\end{figure}

To inspect how the teacher-student interplay evolves, we track the \emph{win rate} of the two views during training, \ie, the fraction of validation positions on which one view ranks the target higher than the other. As shown in Figure~\ref{fig:rankgap}, the teacher rises above the 50\% parity line within the first few hundred steps, so the privileged advantage emerges early and the distillation signal is informative from the beginning. As training proceeds, the student gradually narrows the margin on Video Games and Yelp, reflecting the reachable part of the privileged knowledge being progressively absorbed; on CDs \& Vinyl, the teacher maintains a clear lead throughout, in line with the largest overall gains of PSD on this dataset, where more privileged signal remains to be exploited. We also observe that the teacher's win rate fluctuates sharply in the early stage, \eg, spiking close to 70\% on Yelp before settling down, which corroborates the necessity of the momentum-averaged teacher for smoothing such transient behavior.

\subsubsection{\textbf{Parameter Sensitivity.}}
\begin{figure}[t]
\centering
\includegraphics[width=0.95\linewidth]{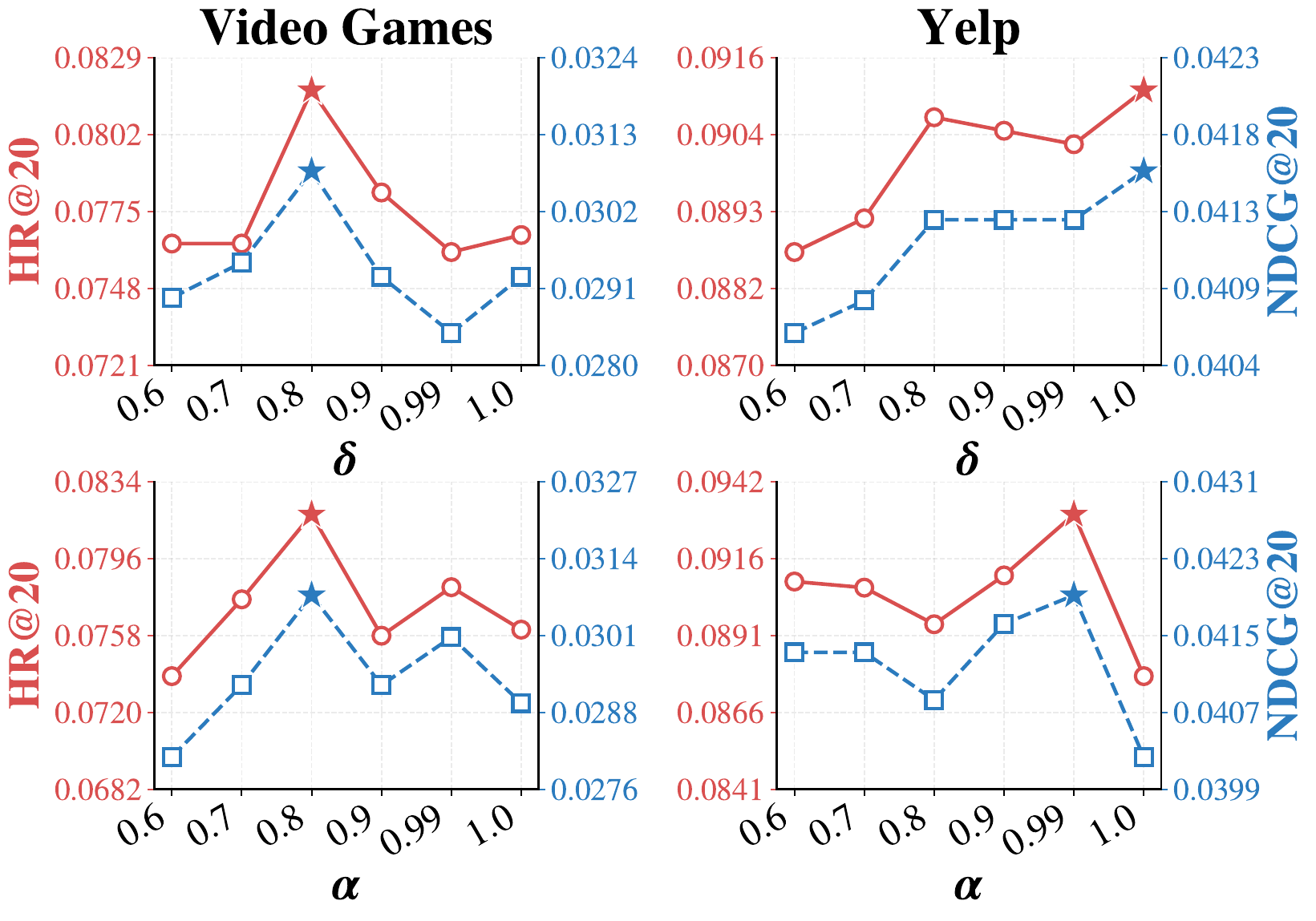}
\vspace{-10pt}
\caption{Sensitivity of PSD to the gate percentile $\delta$ and the EMA rate $\alpha$ on Video Games and Yelp with the SASRec.}
\label{fig:param-sensitivity}
\vspace{-10pt}
\end{figure}

Figure~\ref{fig:param-sensitivity} reports the sensitivity of PSD to its two hyperparameters. \textbf{(i)} For the gate percentile $\delta$, Video Games peaks at $\delta=0.8$, where HR@20 exceeds the ungated variant ($\delta=1.0$) by about 7\%, whereas Yelp varies only mildly across the whole range. This contrast is consistent with Figure~\ref{fig:teacher-student}: the teacher-student gap is wider on Video Games, so filtering unreachable signals pays off more there. \textbf{(ii)} For the EMA rate $\alpha$, moderate to high momentum performs best, with the optimum at $0.8$ on Video Games and $0.99$ on Yelp; pushing $\alpha$ to either extreme is harmful, since a small $\alpha$ lets the target inherit the noise of the instantaneous teacher, while $\alpha=1.0$ freezes the teacher at its initialization. Across the entire grid, PSD remains superior to the backbone, indicating that both hyperparameters are easy to tune in practice.

\section{Related Work}

\subsection{Sequential Recommendation}
Sequential recommendation infers a user's next interaction from the chronological order of her past behaviors. Early studies model item transitions with Markov chains, where FPMC combines matrix factorization with personalized transition matrices~\citep{rendle2010fpmc}. Deep architectures then took over: GRU4Rec models sessions with recurrent networks~\citep{hidasi2016gru4rec}, Caser treats recent behaviors as images for convolution~\citep{tang2018caser}, and self-attention eventually became the dominant choice, with SASRec performing autoregressive next-item prediction~\citep{sasrec} and BERT4Rec reconstructing masked items from bidirectional context~\citep{bert4rec}. Subsequent efforts enrich this paradigm from different angles: S$^3$-Rec~\citep{zhou2020s3rec}, CL4SRec~\citep{cl4srec}, and DuoRec~\citep{duorec} introduce self-supervised or contrastive objectives, FMLP-Rec simplifies the architecture with filter-enhanced MLPs~\citep{zhou2022fmlp}, UniSRec learns transferable representations from item text~\citep{hou2022towards}, and TIGER performs generative retrieval over semantic identifiers~\citep{tiger}. Despite this architectural diversity, these models share the same supervision: a one-hot next-item label conditioned on the prefix, with the logged future left unused. Our work is orthogonal to this line, as it upgrades the supervision itself with training-only future context and applies on top of these backbones without modifying them.

\subsection{Knowledge Distillation}
Knowledge distillation transfers the softened predictions of a teacher into a student~\citep{hinton2015distilling}, later extended to intermediate representations~\citep{romero2015fitnets}. Follow-up studies move beyond model compression: born-again networks show that a student of identical capacity can surpass its teacher~\citep{furlanello2018born}, deep mutual learning trains peer models to teach one another~\citep{zhang2018dml}, and self-distillation turns a single network into its own teacher~\citep{zhang2019byot}. Another line stabilizes a continually updated teacher through parameter averaging, as in Mean Teacher~\citep{tarvainen2017mean}, MoCo~\citep{he2020moco}, and DINO~\citep{caron2021dino}, and recent distillation for large language models further reveals that a static teacher poorly serves an evolving student~\citep{agarwal2024onpolicy,gu2024minillm}. In recommendation, distillation has mainly served ranker compression, from Ranking Distillation~\citep{tang2018ranking} to collaborative and relational variants~\citep{lee2019collaborative,kang2020derrd,kang2021topology}, while PFD~\citep{xu2019privileged} transfers privileged features available only during training and S$^4$Rec~\citep{wei2024leave} and CSRec~\citep{wu2024learning} adopt auxiliary self-distillation to strengthen same-size models. Different from all of them, the teacher in PSD is distinguished by its information set rather than by capacity or extra features: it shares the same parameter with the student, draws its advantage from the future suffix, and transfers only the reachable portion using advantage-reachability privileged information.

\section{Conclusion}
In this paper, we identify the future interactions in logged sequences as training-only privileged information for sequential recommendation, and propose PSD to exploit them without violating the causal serving constraint. PSD evaluates a single backbone under a privileged mask and a causal mask, grounds the teacher on the observed target, and guides the student through gated distillation, where an advantage-reachability gate withholds unreachable dark knowledge and a momentum-averaged teacher stabilizes the self-referential supervision. The whole framework introduces no extra parameters, auxiliary features, or multi-stage training, and leaves the serving cost of the backbone unchanged. Extensive experiments show consistent and substantial gains, and diagnostic analyses verify that the privileged knowledge is genuinely transferred into the causal view. We believe this simple yet effective idea of learning from the future is general enough to benefit a broad range of sequential prediction and decision-making tasks beyond recommendation.

\bibliographystyle{ACM-Reference-Format}
\bibliography{references}

\end{document}